\documentclass[aps,prl,twocolumn,superscriptaddress,citeautoscript,amssymb,reprint,showpacs,floatfix]{revtex4-2}
\usepackage[ansinew]{inputenc}
\usepackage[T1]{fontenc}
\usepackage{float}
\usepackage{amsmath}
\usepackage{ulem}
\usepackage{xurl}
\usepackage{microtype}
\usepackage{mathptmx}
\usepackage{mathrsfs}
\usepackage[english]{babel}
\usepackage{graphicx}
\usepackage{hyperref}
\usepackage[usenames,dvipsnames]{xcolor}
\usepackage{natbib}
\usepackage{makecell}
\usepackage[thinspace,thinqspace]{SIunits}

\usepackage{blindtext}
\usepackage{physics}

\begin{document}
\preprint{APS/123-QED}

\author{Judith Grafenhorst}
\affiliation{Technische Universit\"at Dresden, 01062 Dresden, Germany}

\author{Sarah Krebber}
\affiliation{Kristall- und Materiallabor, Physikalisches Institut, 
Goethe-Universit\"at Frankfurt, 60438 Frankfurt/M, Germany}

\author{Kristin Kliemt}
\affiliation{Kristall- und Materiallabor, Physikalisches Institut, 
Goethe-Universit\"at Frankfurt, 60438 Frankfurt/M, Germany}

\author{Cornelius Krellner}
\affiliation{Kristall- und Materiallabor, Physikalisches Institut, 
Goethe-Universit\"at Frankfurt, 60438 Frankfurt/M, Germany}

\author{Elena Hassinger}
\affiliation{Institute for Quantum Materials and Technologies, Karlsruhe Institute of Technology, 76131 Karlsruhe, Germany}
\affiliation{Technische Universit\"at Dresden, 01062 Dresden, Germany}
\affiliation{Max Planck Institute for Chemical Physics of Solids, 01187 Dresden, Germany}

\author{Ulrike Stockert}
\email{ulrike.stockert@tu-dresden.de}
\affiliation{Technische Universit\"at Dresden, 01062 Dresden, Germany}
\affiliation{Max Planck Institute for Chemical Physics of Solids, 01187 Dresden, Germany}
\affiliation{Institute for Quantum Materials and Technologies, Karlsruhe Institute of Technology, 76131 Karlsruhe, Germany}

\title{Giant thermopower changes related to the resistivity maximum and colossal magnetoresistance in EuCd$_2$P$_2$}
\date{\today}

\begin{abstract}
We present the thermopower of EuCd$_2$P$_2$, a material which exhibits a large resistivity peak with significant magnetic field dependence in the temperature range of 10-25 K. In the same region 
we observe a highly unusual behavior of the thermopower with two sign changes and giant extrema. The overall variation of the thermopower exceeds 4 000 $\upmu$V/K and takes place in an extremely narrow temperature region of less than 5 K. The anomaly is suppressed completely in a small magnetic field of 0.5 T. We discuss this observation using a simple drift-diffusion picture and taking into account that the temperature gradient inducing the thermopower voltage is accompanied by a 
gradient of the electrical resistivity. Our simple estimation yields the correct magnitude, shape, and field dependence of the thermopower anomaly observed in EuCd$_2$P$_2$. These results open a new route to giant thermopower values via gradients of electronic properties. 
\end{abstract}

\maketitle


Thermoelectric materials have a large potential for applications based on the conversion of thermal to electrical energy and reverse \cite{Crane-2025}. The efficiency of these processes is determined by the thermoelectric figure of merit $zT = \sigma S^2 T /\kappa$, which depends on the electrical and thermal conductivities $\sigma$ and $\kappa$, the thermopower $S$, and temperature $T$. A high $zT$ and therefore also a large $S$ is a prerequisite for profitable thermoelectric cooling and appreciable enhancement of the overall power efficiency of technical processes by using waste heat for voltage generation. Beyond that, materials with a large thermopower could be used in thermocouple sensors. A strong field dependence of $S(T)$ would even allow to design thermoelectric devices and sensors controlled by magnetic fields \cite{Wang-2012}. This motivates the exploration of preconditions and mechanisms allowing for large thermopower values and changes. 

The thermopower $S$ of conventional metals and semiconductors contains two contributions, namely from phonon drag $S_\mathrm{drag}$ and charge carrier diffusion $S_\mathrm{diff}$. Large thermopower values exceeding 1 000 $\upmu$V/K and in some cases even 100 000 $\upmu$V/K are usually caused by the phonon drag effect due to momentum transfer from phonons to charge carriers via electron-phonon coupling \cite{Ziman}, e.g. in TiO$_2$ (200 000 $\upmu$V/K) \cite{Thurber-1965, Tang-2009} and Dy$_{1-x}$Sr$_x$MnO$_3$ (220 000 $\upmu$V/K) \cite{Nagaraja-2015}. However, the related good thermal conductivity is detrimental to high values of $zT$ and applications. 
By contrast, the diffusion thermopower $S_\mathrm{diff}$ is typically much smaller, being of the order of 10 $\upmu$V/K for metals and a few 100 $\upmu$V/K for semiconductors below room temperature \cite{Goldsmid}. In a simple picture, it arises from a change of the number and velocity of the mobile charge carriers with temperature $T$ due to the energy dependence of the density of states and the $T$ dependence of the Fermi distribution \cite{Sun-2015}. The resulting imbalance in the presence of a temperature gradient $\nabla T$ leads to diffusion of charge carriers and development of an internal electrical field $E$ limited by a reverse ohmic current $j = \sigma E$. 


Within the semiclassical Boltzman picture, the thermopower of metals and degenerate semiconductors is given by the Mott expression:
\begin{equation}
   S_\mathrm{diff} = \frac{\pi ^2 k_\mathrm{B}^2 T}{3q} \bigg(\frac{d \ln \sigma(\varepsilon)}{d \varepsilon}\bigg)_{E_\mathrm{F}}
\label{Mott}
\end{equation}
This equation relates $S_\mathrm{diff}$ to the logarithmic energy derivative of the conductivity $\sigma(\varepsilon)$ at the Fermi level $E_\mathrm{F}$. $k_\mathrm{B}$ denotes the Boltzmann constant and $q$ the charge of the carriers including sign. Equation \ref{Mott} has the drawback that $\sigma(\varepsilon)$ is not directly accessible by experiments. Therefore, it is often further simplified to estimate the magnitude and temperature dependence of $S$. For simple metals, which can be described as a free electron gas with an energy-independent relaxation time, the derivative of equation \ref{Mott} is of the order of $1/E_\mathrm{F}$ \cite{Paschen-2006}. This yields a linear $T$ dependence $S \propto T/E_\mathrm{F}$ with small absolute values due to the large Fermi energy in normal metals \cite{Behnia-2004, TB-07-6}. By contrast, the thermopower of non-degenerate semiconductors exhibits an inverse temperature proportionality $S \propto E_\mathrm{g}/T$ with the gap width $E_\mathrm{g}$ \cite{Goncalves}. 

These simple concepts cannot be applied to materials with a strong temperature dependence of their electronic properties as for instance charge carrier density, chemical potential, valence, or mobility. In such a situation, any temperature gradient is accompanied by non-negligible gradients of the electronic properties of the material. In this manuscript we demonstrate how this can give rise to giant thermopower values by presenting both experimental results and a simple model taking into account such secondary gradients.
In fact, it has been shown previously that a shift of the chemical potential with $T$ related to a strongly $T$-dependent valence gives rise to an additional thermopower contribution in the intermediate-valence materials EuIr$_2$Si$_2$ and EuNi$_2$P$_2$ \cite{Stockert-2020}. Similarly, the rapidly changing charge carrier mobility with $T$ in Co$_{0.999}$Ni$_{0.001}$Sb$_3$ goes along with a considerable enhancement of $S(T)$ \cite{Sun-2015}. However, in both cases the overall effects have been moderate, leading to absolute thermopower values of at most 50 $\upmu$V/K and 300 $\upmu$V/K, respectively. Here, we investigate $S(T)$ of EuCd$_2$P$_2$, a material with a huge temperature and magnetic field dependence of its electrical resistivity $\rho$ at low $T$ suggesting dramatic changes of the charge carrier mobility and/or number of free charge carriers. In the same regime we observe giant thermopower values in zero magnetic field exceeding those of the above mentioned examples by a factor of more than 10. They can be related directly to $\rho(T)$ based on a simple drift-diffusion picture. Our findings open a new route to giant thermopower values by using gradients of electronic properties. 

EuCd$_2$P$_2$ belongs to a group of trigonal 122 compounds with CaAl$_2$Si$_2$ structure (space group $P\bar{3}m1$) that have been studied in the past as potential thermoelectrics \cite{Artmann-1996, Zhang-2008, Zhang-2010, May-2012, Wubieneh-2016}. More recently, this class of materials -- in particular with composition Eu$T_2 Pn_2$ containing transition metals $T = $ Cd, Zn and pnictides $Pn = $ P, As, Sb -- arouse considerable interest due to the discovery of unconventional electronic and transport properties related to the interplay of band structure topology and magnetism \cite{Kliemt-2025}. The most prominent representative is EuCd$_2$As$_2$ usually classified as a topological Weyl-semimetal \cite{Ma-2019}. Recently, the material has been also suggested as a rare example to host a three dimensional van-Hove singularity \cite{Wu-2024}. Other members of the familiy exhibit a colossal magnetoresistance (CMR) effect, e.g. EuZn$_2$P$_2$ \cite{Krebber-2023}, EuZn$_2$As$_2$ under pressure \cite{Luo-2023}, and the hereinafter investigated material, EuCd$_2$P$_2$ \cite{Wang-2021}.

Magnetism in EuCd$_2$P$_2$ arises from Eu$^{2+}$ moments that order antiferromagnetically at about $T_\mathrm{N} = 11$ K \cite{Schellenberg-2011, Wang-2021, Usachov-2024}. The electrical resistivity exhibits a considerable sample dependence with semiconducting or weakly metallic behavior above about 50 K and a moderate anisotropy for the current within and perpendicular to the $ab$ plane \cite{Wang-2021, Usachov-2024}. The band structure of EuCd$_2$P$_2$ has a gap at the Fermi level as determined consistently by angle-resolved photo-emission spectroscopy (ARPES) experiments \cite{Usachov-2024, Zhang-2023} and band structure calculations \cite{Usachov-2024, Chen-2024a}. Independent of the behavior at higher $T$, a pronounced maximum in $\rho(T)$ is observed between about 10 K and 25 K in zero magnetic field with the resistivity changing by more than one order of magnitude. A moderate magnetic field $B = \mu _0H$ of 1 T is sufficient to suppress the resistivity at the maximum by more than two orders of magnitude with some dependency on crystal quality and growth conditions \cite{Wang-2021, Zhang-2023, Usachov-2024}. The origin of the resistance maximum and CMR effect is a subject of ongoing discussions. It has been shown, that ferromagnetic interactions play an important role \cite{Sunko-2023} and slight modifications in the growth conditions induce a ferromagnetic ground state via charge carrier doping \cite{Chen-2024a}. The most likely scenario appears to be formation and percolation of magnetic polarons \cite{Kopp-2025}. 
Despite these open questions, EuCd$_2$P$_2$ with its huge peak and field dependence of $\rho$ appeared an ideal candidate to study the relation between electrical conductivity $\sigma = 1/\rho$ and thermopower in a situation with an extreme $T$ dependence of the electrical transport properties. Details of our contacting and measurement procedure are given in Supplemental Material \cite{suppl}.





\begin{figure}[t!] 
\centering 
\includegraphics[width=0.95\columnwidth]{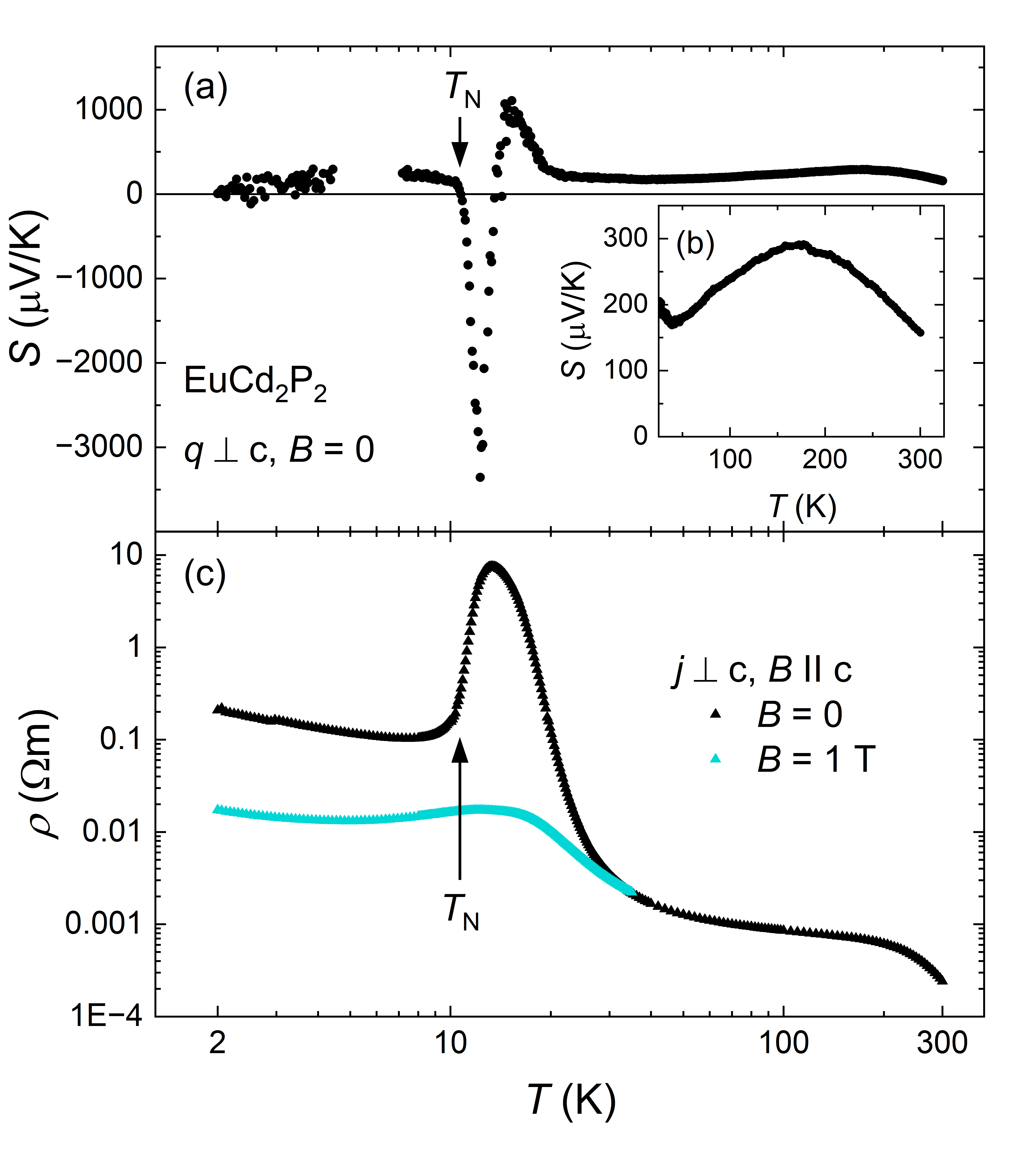}
\caption{(a) Temperature dependence of the thermoelectric power $S$ of EuCd$_2$P$_2$. The antiferromagnetic ordering temperature $T_\mathrm{N}$ is indicated by an arrow. (b) High-$T$ part of $S(T)$ on a larger scale. (c) Electrical resistivity $\rho(T)$ measured on the same sample and contacts in zero magnetic field and $B = \mu_0 H = 1$ T.}
\label{EuCd2P2_S}
\end{figure}

\textit{Results --} The thermopower $S(T)$ of EuCd$_2$P$_2$ is plotted in Fig.~\ref{EuCd2P2_S}a on a logarithmic temperature scale. Fig.~\ref{EuCd2P2_S}b shows the data above 25 K on a linear $T$ scale. In this region EuCd$_2$P$_2$ has a positive thermopower with relatively weak temperature dependence. This is in line with hole-like charge carriers as inferred from Hall effect measurements \cite{Chen-2024a, Sunko-2023}, ARPES \cite{Zhang-2023}, and band structure calculations \cite{Chen-2024a, Usachov-2024}. Below 25 K the thermopower behavior changes dramatically: In a very narrow temperature region between 10 K and 20 K $S(T)$ assumes huge absolute values and switches sign twice. Below 10 K the weak temperature dependence of $S(T)$ with positive values is recovered. Fig.~\ref{EuCd2P2_S}c shows the electrical resistivity 
measured on the same sample and contacts for comparison. In zero magnetic field $\rho(T)$ changes overall by more than four orders of magnitude and exhibits the characteristic maximum around 15 K observed on antiferromagnetic EuCd$_2$P$_2$ samples \cite{Wang-2021, Usachov-2024, Chen-2024a}. The maximum is dramatically suppressed in a relatively small magnetic field of 1 T.

\begin{figure}[t!] 
\centering 
\includegraphics[width=0.95\columnwidth]{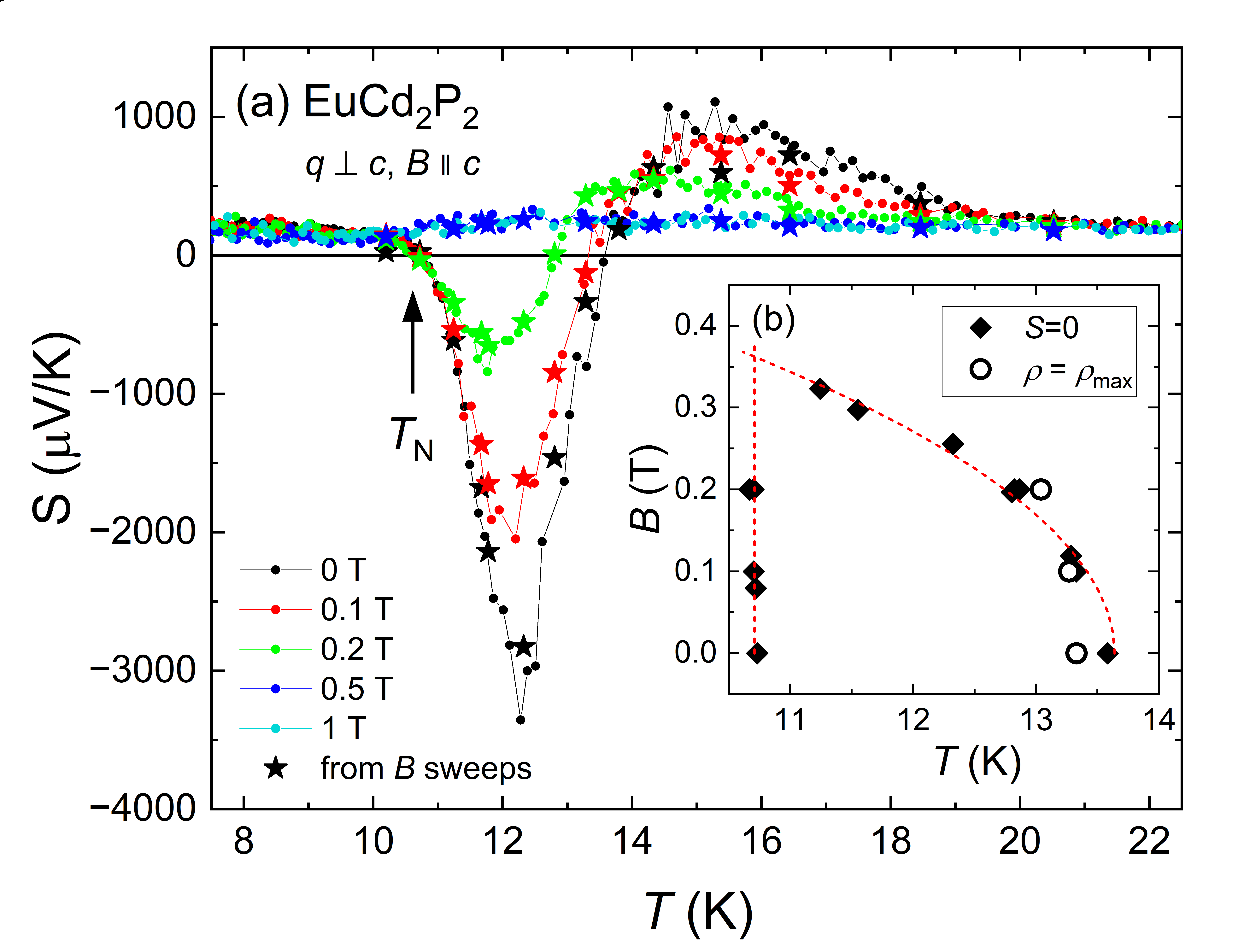}
\caption{(a) Temperature dependence of the thermoelectric power $S$ of EuCd$_2$P$_2$ at different magnetic fields. The results from $T$ sweeps (lines with small symbols) are in good agreement with those from $B$ sweeps (asterisks). (b) Comparison of maxima in $\rho(T)$ and points at which $S$ changes sign in $T$ and $B$ sweeps. The lines are guides to the eye.}
\label{EuCd2P2_SatB}
\end{figure}

The region between 8 K and 22 K is expanded in Fig.~\ref{EuCd2P2_SatB}a. It shows $S(T)$ in zero magnetic field and in magnetic fields $B = \mu_0H$ up to 1 T. In this plot one can trace the giant changes of the thermopower of EuCd$_2$P$_2$ in zero field, which to our knowledge are so far unprecedented concerning the combination of magnitude, sign changes, and extremely narrow $T$ range: Starting from moderate positive values at 10 K the thermopower changes sign upon heating around 10.7~K, goes through a huge minimum at about 12.3 K, switches back to positive values around 13.6~K, reaches a maximum at 15.4~K, and slowly decreases towards higher $T$. That is, in a $T$ range of less than 5 K the thermopower changes sign twice and alters overall by more than 4 000 $\upmu$V/K. The giant slope of $S(T)$ necessitates special measurement conditions as for instance an extremely small temperature difference $\Delta T$ along the sample with $\Delta T / T < 0.12$ \% to determine the intrinsic behavior. 
In addition, the behavior of $S(T)$ has been confirmed on a second sample from a different batch. The respective data are shown in Supplemental Material \cite{suppl}.

\begin{figure}[t!] 
\centering 
\includegraphics[width=0.95\columnwidth]{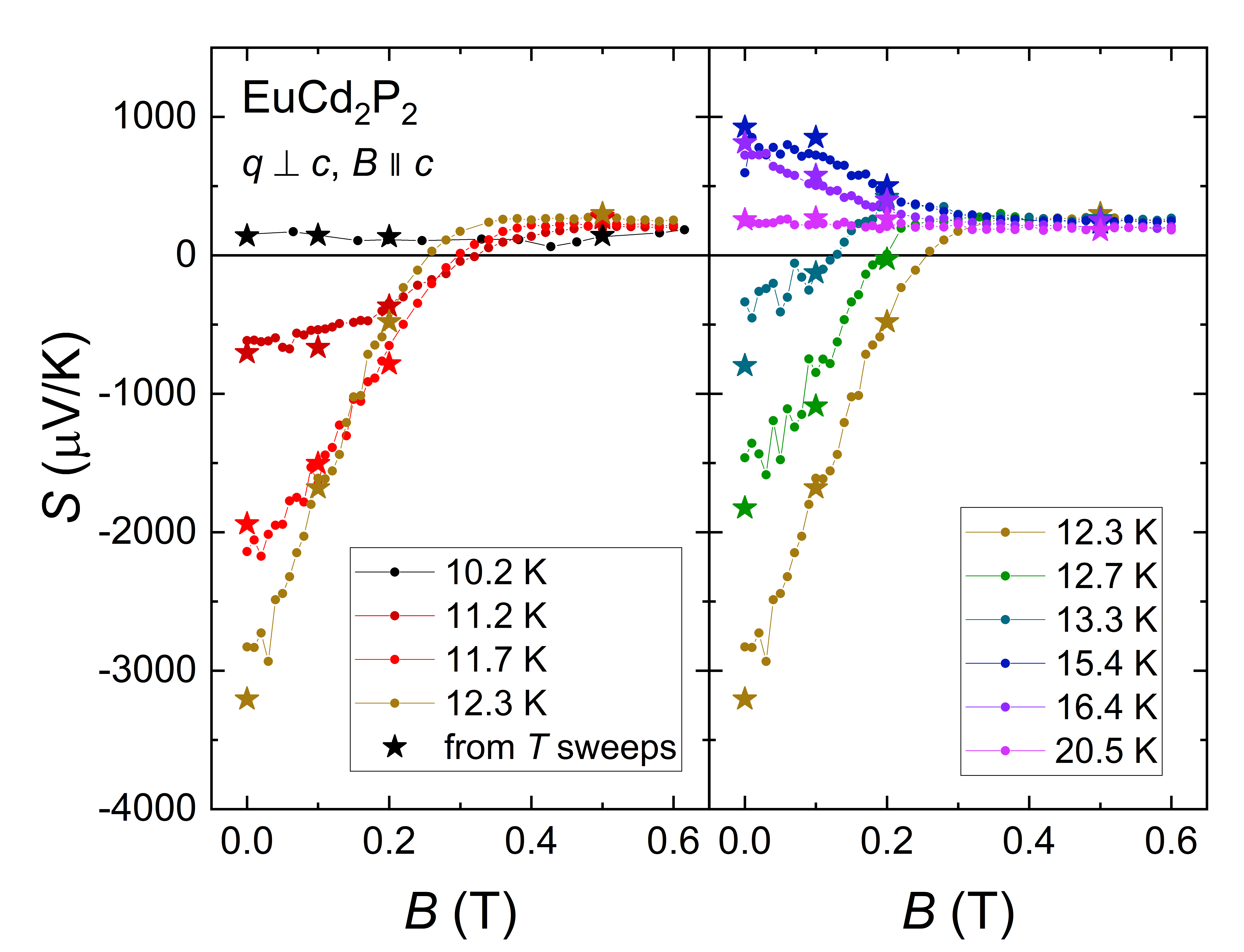}
\caption{Field dependence of the thermoelectric power $S$ of EuCd$_2$P$_2$ at different temperatures. $B$ sweeps (lines with small symbols) and $T$ sweeps (asterisks) exhibit the same behavior.}
\label{EuCd2P2_SatT}
\end{figure}

Application of magnetic fields leads to a rapid suppression of the huge thermopower anomaly between 10 K and 20 K, c.f. Fig.~\ref{EuCd2P2_SatB}a. The strong field dependence is also seen in $B$ sweeps shown in Fig.~\ref{EuCd2P2_SatT}. A field of about 0.3 T is sufficient to eliminate negative thermopower values. In 0.5 T, the anomaly has disappeared completely and the thermopower remains positive without significant $T$ dependence.

Large thermopower values are frequently attributed to phonon drag. However, the complex behaviour of the thermopower anomaly of EuCd$_2$P$_2$ with two sign changes in an extremely narrow $T$ range appears not compatible with dominating phonon drag, although the effect may play some role. We would like to explain this a little more in depth: The drag and diffusion contributions to the thermopower, $S_\mathrm{drag}$ and $S_\mathrm{diff}$, arise from the very same charge carriers, but via two different mechanisms. $S_\mathrm{diff}$ is due to thermal diffusion of the carriers, typically from the hot to the cold end of the sample. In addition, these charge carriers are dragged along by phonons, also usually towards the cold side. Therefore, in simple materials with carriers of a single polarity, both contributions have the same sign leading to a positive thermopower for hole-like charge carriers and a negative one for electrons. A sign change in $S(T)$ as observed for EuCd$_2$P$_2$ could then be caused by a change of the dominating type of charge carriers. This would also imply a polarity reversal of $S_\mathrm{drag}$. However, there is no indication for such an effect in EuCd$_2$P$_2$ neither from ARPES nor band structure calculations, while the Hall effect is dominated by an anomalous contribution. Another mechanism, that may give rise to a change of sign in $S_\mathrm{drag}$, is phonon-charge carrier Umklapp scattering \cite{Belashchenko-1998}. It leads to a reversal of the charge carrier flow and may result in a sign change of $S_\mathrm{drag}$ if strong enough. This effect is largest around $\Theta_D / 5$, $\Theta_D$ being the Debye temperature. It is held responsible for the sign change of $S(T)$ in various materials, e.g. PdCoO$_2$ \cite{Daou-2015} or doped KTaO$_3$ thin films \cite{Nawwar-2025} and might also play a role in semiconductor superlattices \cite{Kubakaddi-1991}. However, for EuCd$_2$P$_2$ the temperature of the minimum in $S(T)$ of about 12 K appears rather low compared to the estimate of $\Theta_D = 147$ K from specific heat \cite{Usachov-2024}. Most important, to the best of our knowledge there is no single mechanism that gives rise to two sign changes in $S_\mathrm{drag}$. The only conceivable situations would be a combination of effects or switching the carrier type twice. This appears highly unlikely and practically rules out phonon drag as the main origin for the rather complex and extremely narrow anomaly in $S(T)$.

Before discussing an alternative scenario for the unusual thermopower of EuCd$_2$P$_2$ we would like to rank the observed values and field dependence. In literature the terms "giant" or "colossal" thermopower have been used for a wide range of absolute values. Some of the largest thermopowers in bulk materials have been observed in manganates as Dy$_{1-x}$Sr$_x$MnO$_3$ (220 000 $\upmu$V/K) \cite{Nagaraja-2015} and Gd$_{1-x}$Sr$_x$MnO$_3$ (35 000 $\upmu$V/K) \cite{Sagar-2010}, in TiO$_2$ (200 000 $\upmu$V/K) \cite{Thurber-1965}, and in FeSb$_2$ (45 000 $\upmu$V/K) \cite{Bentien-2007}. However, as mentioned in our introduction the thermopower in these systems is at least to some extend enhanced by phonon drag \cite{Nagaraja-2015, Tang-2009, Battiato-2015}. By contrast, the diffusion thermopower in metals and semiconductors is typically three orders of magnitude lower with values of 10-100~$\upmu$V/K \cite{Goldsmid}. Based on this scale, the thermopower of 400 $\upmu$V/K observed in the clathrate system Ba$_8$(Cu,Zn)$_x$Ge$_{46-x}$ has also been labeled as "giant" \cite{Bednar-2010}. The observation of absolute values exceeding 3 000 $\upmu$V/K in EuCd$_2$P$_2$ is therefore clearly remarkable.

\begin{table}
\caption{\label{magnetothermopower}%
Materials claimed to exhibit a giant magnetothermopower. The maximum reported value d$S_\mathrm{max} = | S(B)-S(0) |$ is given together with the corresponding temperature $T(\mathrm{d}S_\mathrm{max})$ and field $B(\mathrm{d}S_\mathrm{max})$, at which it was determined. In addition, the steepest slope |d$S$/d$B$|$_\mathrm{max}$ is specified.}
\begin{ruledtabular}
\begin{tabular}{cccccc}
    Material & d$S_\mathrm{max}$ & $T(\mathrm{d}S_\mathrm{max})$ & $B(\mathrm{d}S_\mathrm{max})$ & |d$S$/d$B$|$_\mathrm{max}$ & reference \\
     & [$\upmu$V/K] & [K] & [T] & [$\upmu$V/KT] & \\
    \colrule
    EuCd$_2$P$_2$ & 3 600 & 12 & 0.5 & 13 000 & this study \\
    \colrule
    InAs & 10 000 & 8 & 29 & 800 & \cite{Jaoui-2020} \\
    Bi & 4 000 & 7 & 5 & 1 500 & \cite{Mangez-1976} \\
    NbP & 800 & 50 & 9 & 150 & \cite{Stockert-2017} \\
    Ag$_{2-\delta}$Te & 470 & 110 & 7 & 150 & \cite{Sun-2003} \\
\end{tabular}
\end{ruledtabular}
\end{table}

Besides the large absolute values of $S$ observed in EuCd$_2$P$_2$ in zero magnetic field, the dramatic suppression in relatively small magnetic fields is also remarkable. Typically "giant" magnetothermopower values defined as $| S(B)-S(0) |$ are claimed for applied fields of several tesla. Some examples are summarized in table \ref{magnetothermopower} in comparison to EuCd$_2$P$_2$. Besides the maximum change of $S$ in magnetic fields, also the steepest slope |d$S$/d$B$|$_\mathrm{max}$ is given. The values in table \ref{magnetothermopower} demonstrate that EuCd$_2$P$_2$ is indeed exceptional, both concerning the low field required to induce a large magnetothermopower and the huge slope of $S(B)$. Interestingly, the absolute values of $S(T)$ of EuCd$_2$P$_2$ decrease with increasing magnetic field, while all other materials mentioned in table \ref{magnetothermopower} exhibit an opposite field dependence. 

Summing up, the thermopower of EuCd$_2$P$_2$ in the $T$ range of 10-20 K is highly unconventional with respect to four facts: (1) huge absolute values primarily not due to phonon drag, (2) an extremely strong temperature dependence with the slope exceeding 2 000 $\upmu$V/K$^2$, (3) a dramatic suppression of $S$ in exceptionally small magnetic fields, and (4) an overall unusual $T$ dependence with two sign changes within 3 K.

\textit{Discussion --} Since the thermopower anomaly of EuCd$_2$P$_2$ is not due to phonon drag, at least not exclusively, we have to search for an alternative explanation. Comparing the thermopower to the electrical resistivity reveals that both properties exhibit highly unconventional behavior in a very similar temperature and field range.  
It appears most natural that both effects should be related: The giant $T$ dependence of $\rho$ -- changing for instance by more than an order of magnitude upon cooling from 20 K to 15 K -- leads to a sizeable resistivity gradient in the presence of a temperature gradient. This in turn is expected to have an impact on the diffusion thermopower. Consistently, the maximum in $\rho(T)$ at 13.5 K coincides with a sign change and, therefore, also small values of $S(T)$. This is demonstrated in Fig.~\ref{EuCd2P2_SatB}b. It shows the positions of the maxima in $\rho(T)$ and where $S = 0$. The second sign change in $S(T)$ at lower $T$ occurs close to the ordering temperature $T_{\mathrm{N}}$. It has bee determined  from specific heat measurements on samples from the same batch to be $T_\mathrm{N} = 10.6$ K in zero magnetic field with an only weak field dependence up to 0.4 T \cite{Usachov-2024}.

The relation between resistivity and thermopower anomalies in EuCd$_2$P$_2$ cannot be understood from simple diffusion pictures for normal metals or semiconductors: For example, around 11~K, i.e. close to the largest slope of $\rho(T)$ a temperature difference of 1 \% gives rise to more than 20 \% change in resistivity. This effect is much stronger than, e.g., thermal broadening of the Fermi-Dirac distribution, which is the main source for the thermopower in simple models as the Mott relation. 

\begin{figure}[t!] 
\centering 
\includegraphics[width=0.95\columnwidth]{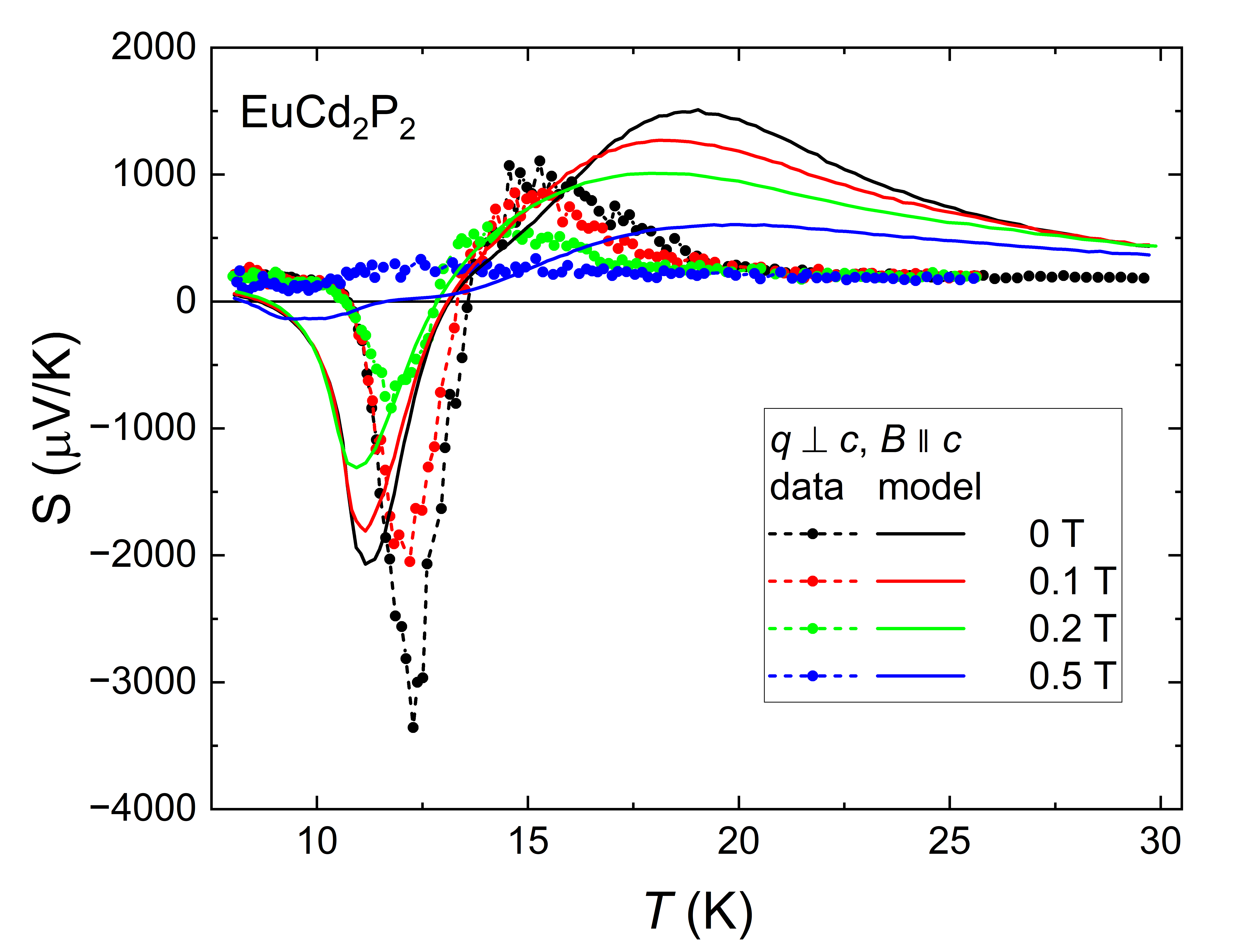}
\caption{Temperature dependence of the thermoelectric power $S$ of EuCd$_2$P$_2$ at different magnetic fields. Lines are calculations for a drift-diffusion model as explained in the main text.}
\label{EuCd2P2_model}
\end{figure}

In order to evaluate the influence of the resistivity anomalies of EuCd$_2$P$_2$ on the diffusion thermopower we start from a macroscopic transport picture. At this point, we are only interested in the impact of the strong changes in $\rho(T)$ on $S_\mathrm{diff}$. The material is treated as being single phase with homogeneous properties apart from its strongly $T$-dependent electrical transport. That is, we do not consider local inhomogeneities arising from ferromagnetic-cluster or magnetic-polaron formation. In this picture the thermopower results from a balance between two currents: a diffusion current $j_\mathrm{diff}$ due to a change of electronic and transport properties with $T$ and a drift current $j_\mathrm{drift}$ arising from the developing electrical field $E$. In our evaluation of the thermopower we allow explicitly for a variation in the charge carrier density $n$ and diffusion coefficient $D$ along the sample. The latter is usually assumed constant or quasi-constant in literature, which is an important difference to our calculation. The  situation with variable $n$ and $D$ is described by the Stratton equation \cite{Stratton-1962}. The total electrical current along a specific direction (e.g. $x$) amounts to
\begin{equation}
    j_x = j_\mathrm{diff} + j_\mathrm{drift} = \sigma E_x-q\frac{\partial (nD)}{\partial x}
\label{stratton}
\end{equation}
For simplicity we ignore anisotropies of transport coefficients and omit the corresponding subscripts. In thermopower measurements, the net current is zero and the diffusion and drift currents have opposite direction: $j_\mathrm{diff} = - j_\mathrm{drift}$. Assuming a constant temperature gradient along the sample we may replace the derivative with respect to $x$ by the one with respect to $T$: $\partial (nD)/\partial x = \partial (nD)/ \partial T \cdot \nabla T$. In addition we use the relations between diffusion coefficient $D$, mobility $\mu$, and electrical conductivity $\sigma$ to replace $D = \mu k_\mathrm{B} T/q$ by $\sigma = n \mu q$, with the Boltzmann constant $k_\mathrm{B}$. Thus, we get a direct relation between the diffusion thermopower and electrical conductivity:
\begin{equation}
    S_\mathrm{diff} = \frac{E}{\nabla T} = \frac{k_\mathrm{B}}{q} \frac{1}{\sigma} \frac{\partial (\sigma T)}{\partial T}=\frac{k_\mathrm{B}}{q} \Big( 1+\frac{\partial \ln \sigma}{\partial \ln T} \Big)
\label{SvsSigma}
\end{equation}
In a trigonal system as EuCd$_2$P$_2$, electrical conductivity and resistivity along high-symmetry directions are directly inverse, i.e., for $j$ along [210] we can take $\sigma = 1/ \rho$. Therefore, equation \ref{SvsSigma} allows a calculation of the diffufion thermopower from the electrical resistivity. The results obtained for EuCd$_2$P$_2$ and hole-like charge carriers are plotted in Fig.~\ref{EuCd2P2_model}. The resemblance of calculation and data is extremely good taking into account the simplicity of our assumptions. In fact, equation \ref{SvsSigma} predicts the correct shape, sign, magnitude, and field dependence of the thermopower. The main difference is the width of the anomalies in $S(T)$. The strongest temperature dependence is observed in a more narrow regime than predicted by equation \ref{SvsSigma}. These deviations can be attributed to the simplicity of our drift-diffusion picture of classical particles, which does not account for impact from different scattering or transport channels, local inhomogeneities due to ferromagnetic domain or polaron formation, percolation effects, and variations in the chemical potential with $T$. The relevance of these factors might be evaluated by systematically studying the thermopower of crystals grown under different conditions. In fact, it has been shown, that small variations in the charge carrier concentration may lead to dramatic changes in the magnetic and transport properties of the material \cite{Chen-2024a}. Moreover, we cannot exclude that phonon drag plays a certain role in the material and gives rise to an additional contribution to $S(T)$.  

Altogether, our very simple model is able to account for all four of the above-mentioned aspects of the highly unconventional thermopower of EuCd$_2$P$_2$. In particular we would like to emphasize that there is no free parameter in our calculation: The only ingredients are the electrical resistivity and the charge carrier type, while the microscopic origin of the strong $T$ dependence of $\rho (T)$ and the CMR are not considered. Therefore, the similarity between the calculated and measured thermopower curves is indeed impressive. It clearly demonstrates, that the unusual temperature and field dependencies of $S$ in EuCd$_2$P$_2$ are caused by the strong changes of the electrical resistivity with $T$ and $B$. 

Despite the large thermopower observed in EuCd$_2$P$_2$, the thermoelectric figure of merit $zT$ is very small -- about $10^{-7}$ at 12 K. This is mainly due to the extremely high electrical resistivity of the material. In general, usage of the presented effect in thermoelectric devices is limited by the small temperature region of the thermopower enhancement. It is directly related to the strong $T$ dependence of the electrical resistivity, which cannot be perpetuated over a large $T$ range. This drawback might be overcome to some extend by material's customization to operating environments. 
Independent of these limitations, our finding proofs the direct relation between thermopower and resistivity anomalies in EuCd$_2$P$_2$. This shows that in general, it is possible to generate a large thermopower as needed for decent $zT$ values via secondary gradients of electronic properties. A possible direction for exploitation of this mechanism is the search for materials with secondary gradients of other electronic properties or the use of internal compositional gradients induced via position-dependent doping.


\textit{Data Availability Statement}
 The data that support the findings of this article are openly
 available \cite{data}.

\textit{Acknowledgments --} We would like to thank B. Fauqu\'e and K. Behnia  for clarifying discussions. We acknowledge funding from the Deutsche Forschungsgemeinschaft (DFG, German Research Foundation) through the Collaborative Research Centers CRC 288 (422213477, Project No. A03) and CRC 1143 (247310070, Project No. C10) and through the W\"urzburg-Dresden Cluster of Excellence ctd.qmat - Complexity, Topology and Dynamics in Quantum Matter (EXC 2147, project-id 390858490). We acknowledge financial support from the Max Planck Society through the Physics of Quantum Materials department and the research group ''Physics of
Unconventional Metals and Superconductors (PUMAS)''.


\bibliographystyle{apsrev4-2}
\bibliography{EuCd2P2-bib}




\clearpage

\pagebreak
\begin{center}
\textbf{\large Supplemental Material}
\end{center}

\setcounter{equation}{0}
\setcounter{figure}{0}
\setcounter{table}{0}
\setcounter{page}{1}
\makeatletter
\renewcommand{\theequation}{S\arabic{equation}}
\renewcommand{\thefigure}{S\arabic{figure}}



\section{S1: Experimental Techniques}

Thermal and electrical transport measurements were performed on single crystals of EuCd$_2$P$_2$ grown from Sn flux \cite{Usachov-2024}. The crystals grow as plates with the short dimension along [001]. They were cut to bars with the long extension along [210]. Four micro-contacts were sputtered onto the sample consisting of a 10 nm titanium layer covered by a 150 nm gold layer. Typical sample dimensions were $3 \times 1 \ \times 0.5$ mm$^3$ with the distance between the middle contacts being about 1.5 mm. 

Thermoelectric transport measurements were performed in a commercial Physical Property Measurement System (PPMS) from Quantum Design (QD) using the thermal transport option (TTO). For this purpose, Au-coated Cu bars were attached with silver paint to the microcontacts. The TTO allows measurement of thermal conductivity, thermoelectric power, and AC electrical resistivity in a single run and using the same contacts. However, in order to avoid sample heating and long relaxation times between thermal measurements, we performed resistivity measurement in separate runs. 

The thermal conductivity $\kappa$ and thermopower $S$ were measured using a standard steady-state method with a two-thermometer-one-heater configuration in the $T$ range of 2 K - 300 K. The heat current $j_Q$ was applied along [210] and the magnetic field along [001]. For temperature-dependent measurements, square-wave heat pulses were applied and the time evolution of temperature difference and voltage along the sample monitored, while continuously sweeping the bath temperature. The thermal conductivity and thermopower were then calculated by a fitting routine, which also estimates the adequate power and duration for the next heat pulse. Magnetic-field-dependent measurements were performed at constant bath temperature and field using a constant heater power. Due to the negligible field-dependence of the thermal conductivity, it is assumed that measurements effectively take place at constant average sample $T$. Several data points were taken at each field to allow for averaging before sweeping to the next set point. In addition, continuous measurements were performed while sweeping the field in regimes with very week field-dependence. These measurements are much faster. However, data spacing is much larger due to the minimum sweep rate of the magnet.

\begin{figure} 
\centering 
\includegraphics[width=0.95\columnwidth]{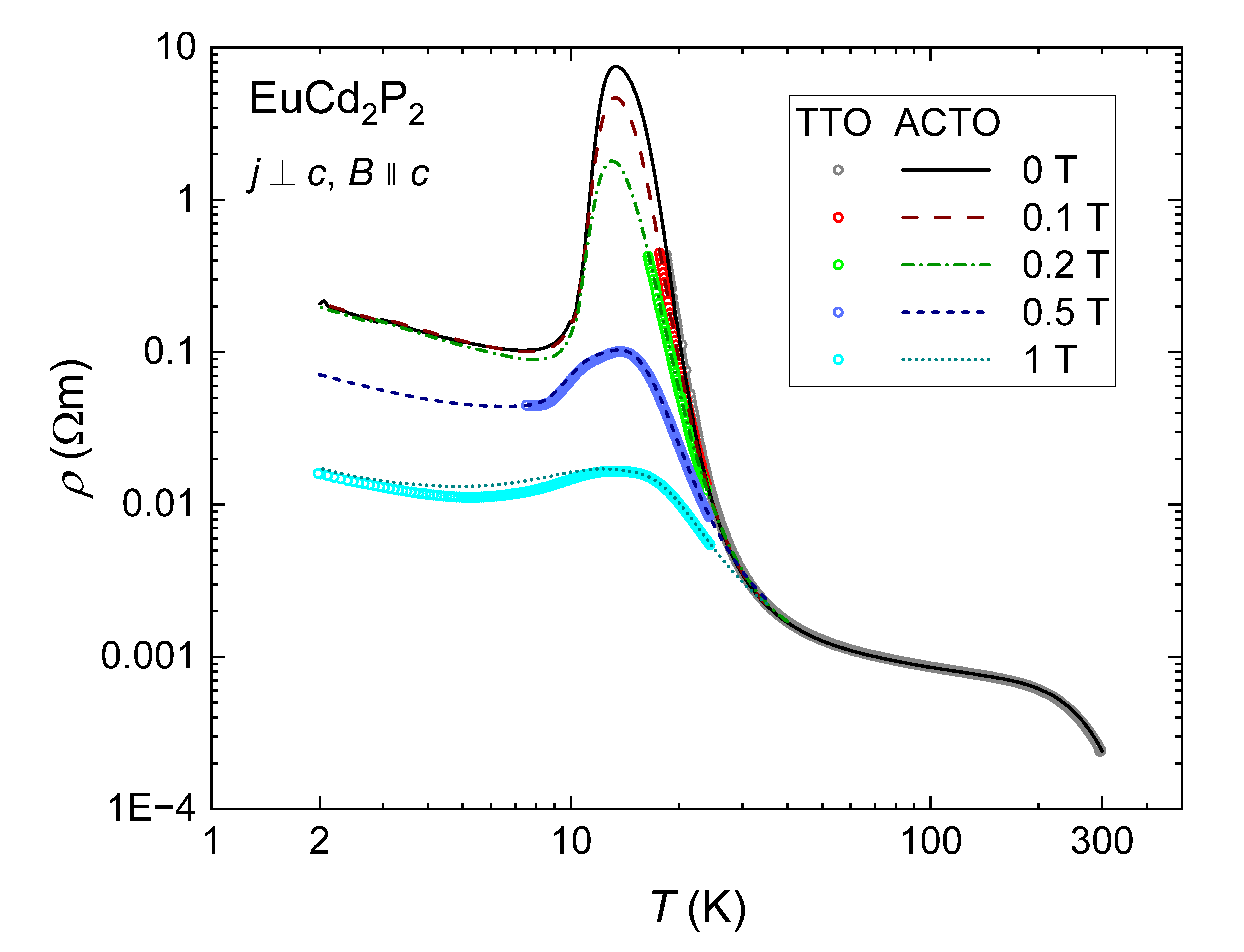}
\caption{Comparison of the measured temperature dependence of the electrical resistivity $\rho$ of EuCd$_2$P$_2$ obtained on the same contacts, but with two different measurement techniques. Data points have been measured using the TTO. The lines correspond to measurements with the ACTO. Both methods yield the same results at high $T$ and $B$. However, the TTO was not able to resolve very large resistances, as observed in low fields below 20 K in EuCd$_2$P$_2$.}
\label{EuCd2P2_rho}
\end{figure}

The TTO measures the electrical resistivity $\rho$ using a 4-point AC technique. However, $\rho$ could not be resolved at low $T$ and $B$ probably because of sample self heating. Therefore, we performed additional measurements using the AC transport option (ACTO) of the PPMS and a low excitation current of 0.01 mA. For this purpose, we attached Au wires to the same micro-contacts used for thermoelectric transport measurements. The sample had a large contact area to the bath, and measuerments were performed in He atmosphere for thermalization. Both methods, TTO and ACTO, are in good agreement confirming the equal geometry factor and the quality of our thermal transport contacts. The results are shown in Fig.~\ref{EuCd2P2_rho}.

In the $T$ range between 10 K and 20 K, extremely large values of $\rho$ and a giant $T$ dependence of both $\rho$ and $S$ are observed in EuCd$_2$P$_2$. For instance, changing the temperature from 11.5 K to 11.6 K, i.e. by less than 1 \%, leads to an increase of $\rho$ by 22 \%. In the same $T$ range, $S$ changes by about 260 $\upmu$V/K corresponding to 17 \% at this temperature. In order to ensure reliability of our data, we took several precautions:

\begin{itemize}
    \item The dramatic temperature dependence of $S$ requires measurements at very low $T$ gradient to establish a quasi-constant situation along the sample. Below 20 K we applied an extremely small $T$ difference $\Delta T$ along the sample with $\Delta T /T < 0.12$ \%. This tiny gradient is responsible for the enhanced noise at low $T$ in our data. 
    \item We confirmed that our results from $T$ sweeps and $B$ sweeps are consistent. This is important, because the $T$ dependent thermopower was measured while sweeping the bath temperature. The agreement with results from field sweeps validates, that the sweep rate of 0.1 K/min was sufficiently low.
    \item In addition, we confirmed that $S$ is roughly independent of the applied temperature gradient by measuring exemplarily field sweeps at 11.5 K for two different values of $\Delta T$, namely 5 mK and 7.5 mK. The resulting thermopower curves can be scaled by a factor of 1.1, which is small compared to the observed field and temperature dependencies.
\end{itemize}

\section{S2: Thermal conductivity} 

\begin{figure} 
\centering 
\includegraphics[width=0.95\columnwidth]{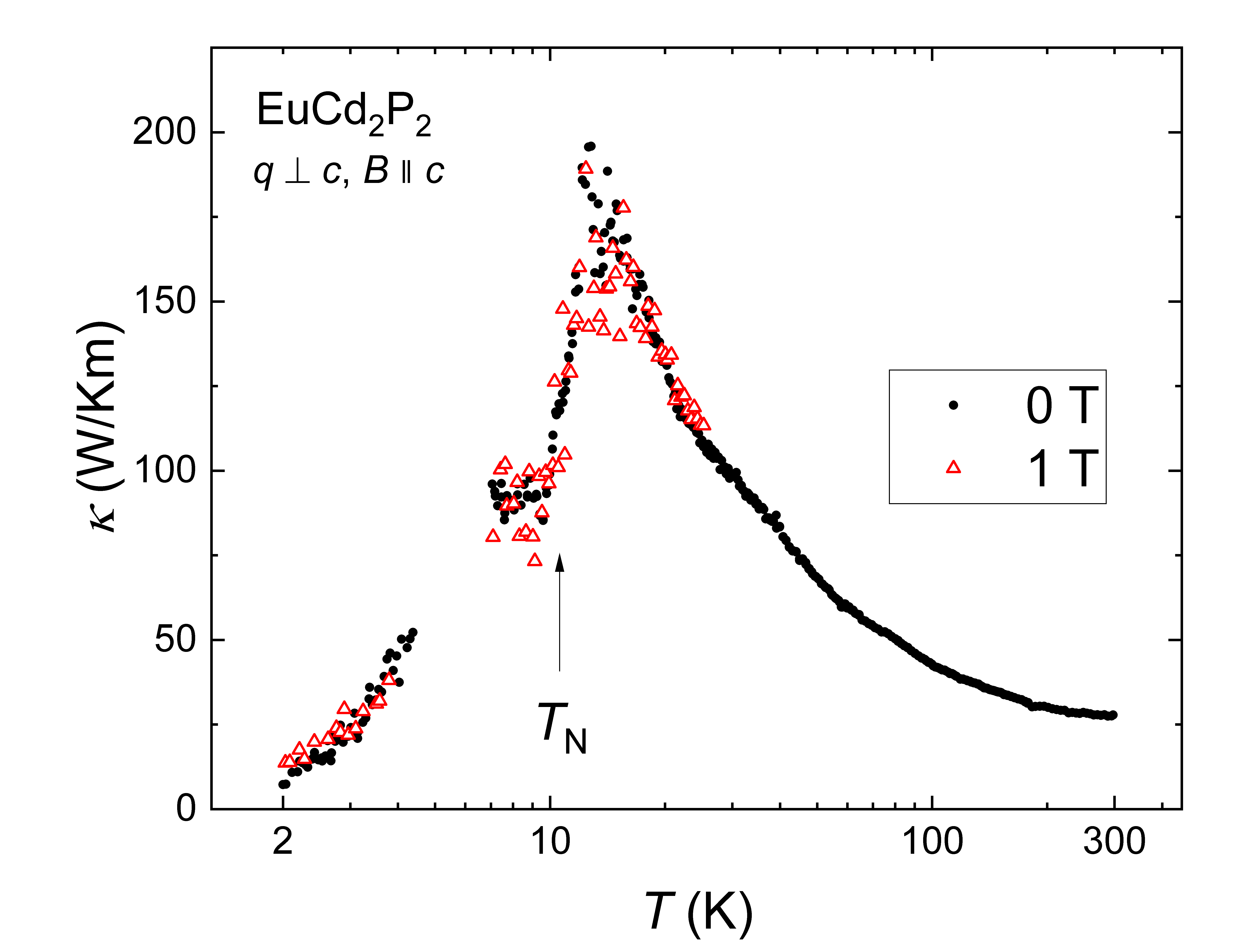}
\caption{Temperature dependence of the thermal conductivity $\kappa$ of EuCd$_2$P$_2$ in zero magnetic field and a field of 1 T.}
\label{EuCd2P2_kappa}
\end{figure}

\begin{figure} 
\centering 
\includegraphics[width=0.95\columnwidth]{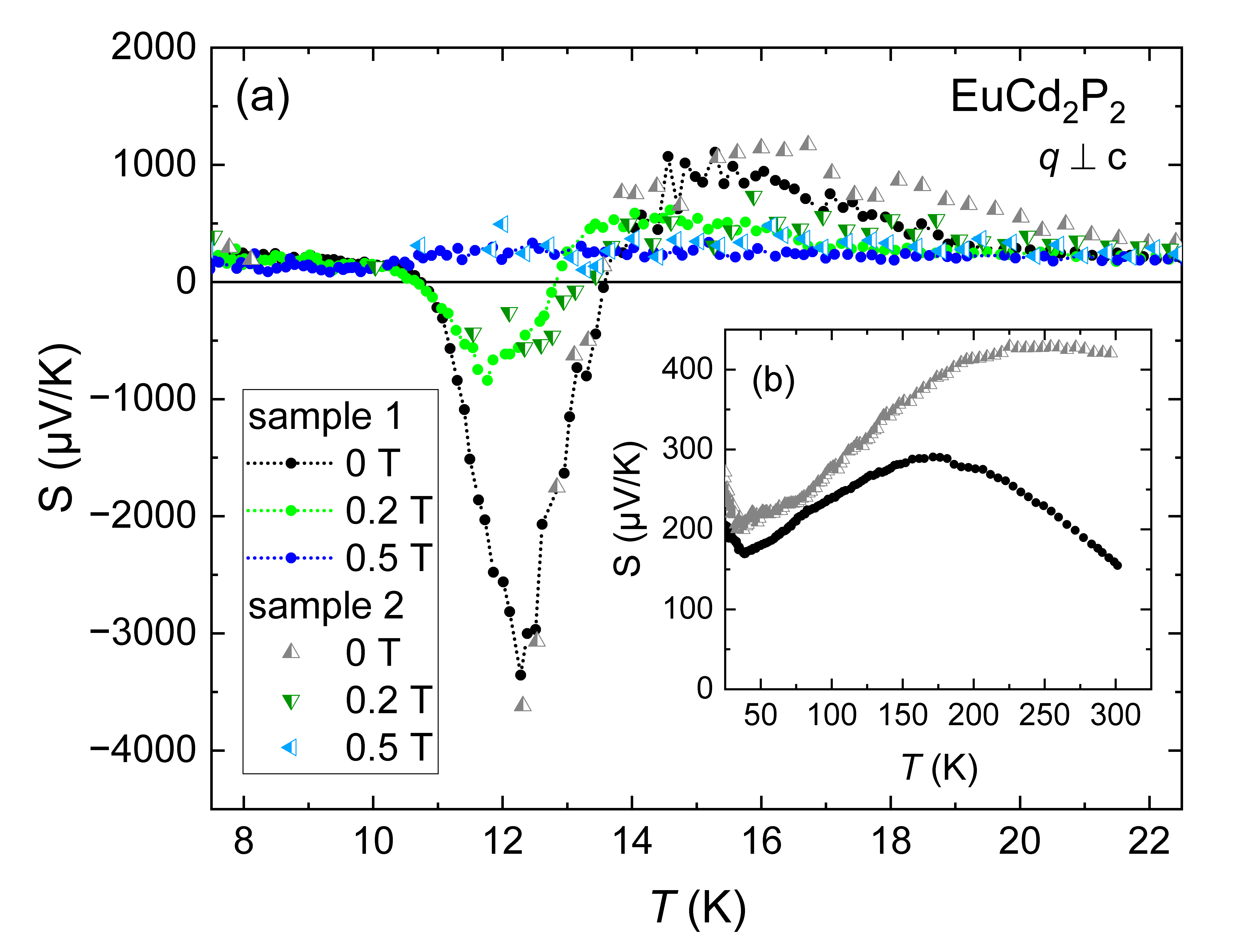}
\caption{Comparison of the thermopower for two different samples of EuCd$_2$P$_2$. (a) Anomaly at low $T$ in different magnetic fields. (b) Behavior of the thermopower in the regime 25 K - 300 K.}
\label{EuCd2P2_SvsT_sample2}
\end{figure}

Fig.~\ref{EuCd2P2_kappa} shows the thermal conductivity $\kappa$ of EuCd$_2$P$_2$ in zero magnetic field and 1 T. The rather large noise below 20 K is due to the very small $T$ gradient applied in this $T$ range as explained above. In fact, in the temperature range between about 4 K and 7 K $\kappa$ could not be resolved due to the small $T$ gradient and large thermal conductivity of the sample. Apart from this region, EuCd$_2$P$_2$ exhibits the typical behavior of a semiconductor with a thermal conductivity due to phonons: $\kappa(T)$ initially increases with increasing $T$, reaches a maximum around 15 K and decreases towards higher $T$. The electronic contribution to the thermal conductivity estimated from the electrical resistivity using the Wiedemann-Franz law is more than six orders below the measured $\kappa$ values. This clearly shows, that $\kappa$ is solely due to phonons. Application of a magnetic field does not alter the thermal conductivity, which is also in line with a purely phononic $\kappa$.

\section{S3: Thermopower of a second sample}
In order to confirm the observed behavior and absolute values of the thermopower of EuCd$_2$P$_2$ we studied a second sample (sample 2) from a different batch. The results are compared to those presented in the main text (sample 1) in Fig.~\ref{EuCd2P2_SvsT_sample2}. The dimensions of sample 2 were slightly shorter than for sample 1. Therefore, the corresponding measurement was more noisy, and we had problems to resolve the temperature gradient below 12 K. However, as demonstrated in Fig.~\ref{EuCd2P2_SvsT_sample2}a, both samples exhibit a very similar temperature and field dependence of the anomaly at low $T$. Above 25 K the thermopower curves of both samples differ slightly, cf. Fig.~\ref{EuCd2P2_SvsT_sample2}b. This may be caused by a different amount impurities acting as dopants.


\end{document}